# Uncovering structure-property relationships of materials by subgroup discovery

Bryan R. Goldsmith*,l,1, Mario Boley[l,1,2], Jilles Vreeken[2], Matthias Scheffler[1], and Luca M. Ghiringhelli*,1

[1]Fritz-Haber-Institut der Max-Planck-Gesellschaft, Faradayweg 4-6, D-14195 Berlin, Germany
[2]Max Planck Institute for Informatics, Campus Mitte, 66123 Saarbrücken, Germany

**E-mails**: *goldsmith@fhi-berlin.mpg.de; ghiringhelli@fhi-berlin.mpg.de

**Abstract**

Subgroup discovery (SGD) is presented here as a data-mining approach to help find interpretable local patterns, correlations, and descriptors of a target property in materials-science data. Specifically, we will be concerned with data generated by density-functional theory calculations. At first, we demonstrate that SGD can identify physically meaningful models that classify the crystal structures of 82 octet binary semiconductors as either rocksalt or zincblende. SGD identifies an interpretable two-dimensional model derived from only the atomic radii of valence $s$ and $p$ orbitals that properly classifies the crystal structures for 79 of the 82 octet binary semiconductors. The SGD framework is subsequently applied to 24 400 configurations of neutral gas-phase gold clusters with 5 to 14 atoms to discern general patterns between geometrical and physicochemical properties. For example, SGD helps find that van der Waals interactions within gold clusters are linearly correlated with their radius of gyration and are weaker for planar clusters than for nonplanar clusters. Also, a descriptor that predicts a local linear correlation between the chemical hardness and the cluster isomer stability is found for the even-sized gold clusters.

## 1. Introduction

Rational design of advanced functional materials, e.g., active and selective catalysts [1], efficient thermoelectrics [2], and high-temperature superconductors [3], requires an understanding of the underlying fundamental physical mechanisms. Identifying interpretable [4] rule-based models that describe materials phenomena is therefore critical. For example, Brønsted-Evans-Polanyi relations allow for an efficient approach to estimate activation energies of similar reactions [5], the Goldschmidt tolerance factor is an indicator for the stability and structure of ionic crystals [6,7], and the thermoelectric figure of merit, as well as semi-empirical models, guide the design of thermoelectrics [8]. However, in general it remains difficult to extract insights from materials-science data and to discover rules for desired materials properties and function.

Big-data analytics tools, e.g., statistical/machine learning, compressed sensing, and data mining methods, are becoming widely applied in the materials-science community [9-19]. Efficient algorithms for model selection can be used for the estimation of alloy formation energies [20], and machine-learning algorithms trained on reaction data can help predict the crystallization outcomes of materials [21]. Data analytics tools can identify descriptors [22,23] that characterize properties such as hole traps in amorphous silicon [24] and the intrinsic dielectric breakdown field of insulators [25]. Importantly, the application of big-data analytics to obtain material insights and to predict novel materials can be enhanced by the availability of large materials repositories, e.g., AFLOWLIB, Computational Materials Repository, Electronic Structure Project, Materials Project, Novel Materials Discovery (NOMAD), Open Quantum Materials Database (OQMD), and Pauling file [26]. Our objective is to develop and exploit big-data analytics tools to discover materials insights and to predict advanced materials from large collections of materials data stored within the NOMAD Archive [27].

Big-data analytics applied to materials-science data often focuses on the inference of a global prediction model for some property of interest for a given class of materials. However, the underlying mechanism for some target property could differ for different materials within a large pool of materials-science data. Consequently, a global model fitted to the entire dataset may be difficult to interpret and may well hide or incorrectly describe the actuating physical mechanisms [28]. In these situations, local models describing subgroups would be advantageous to global models. For illustration (see Figure 1), a globally optimal regression model could predict a negative relationship between two material properties, whereas among subgroups there exists a positive relationship. As a more physical example, the transition metals of the periodic table are a subgroup, and the actinides, lanthanides, and halogens are other subgroups. Thus, identification of subgroups is useful to gain an understanding of similarities and differences between materials.

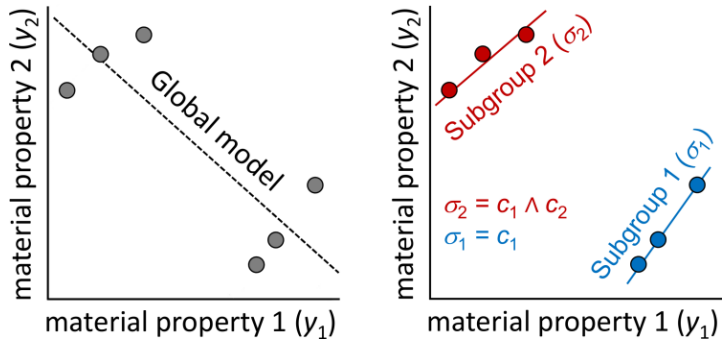

**Figure 1.** A schematic of an optimal global regression model predicting a negative relationship between material properties $y_1$ and $y_2$ (with non-zero error), whereas two subpopulations allow a nearly perfect fit using a linear regression function with positive slope. Subgroup discovery aims to describe such subpopulations by Boolean selector functions ($\sigma_1$ and $\sigma_2$) defined as conjunctions (logical AND, denoted as ∧) of basic selectors (the $c_i$).

In this paper we demonstrate a multipurpose data-mining algorithm called subgroup discovery (SGD) to identify and describe local patterns, correlations, and descriptors in materials-science data according to some desired target property (or properties) [29-32]. At first, we begin by formulating the subgroup discovery algorithm for materials-science applications. Next, we demonstrate that SGD can identify physically meaningful models that classify the crystal structures of 82 octet binary semiconductors as either rocksalt (RS) or zincblende (ZB) from only information of its chemical composition. The octet binary compounds have long been studied [22,33-40], and we consider it an exemplary dataset for the demonstration of SGD to find descriptors of materials. Notably, SGD helps us to find a two-dimensional model derived from only the atomic radii of valence $s$ and $p$ orbitals that properly classify the crystal structures for 79 of the 82 octet binary semiconductors. Subsequently, we apply SGD to 24 400 configurations of neutral gas-phase gold clusters with 5 to 14 atoms. Small gold clusters have different physical and chemical properties than their bulk counterpart, and they exhibit a diverse array of physicochemical properties depending on their size and shape [41-50]. The aim of investigating gold clusters here is two-fold: (1) to search for general structure-property relationships holding across gold clusters of different sizes and vastly different configurations; and (2) to demonstrate the versatility of subgroup discovery on a large and heterogeneous dataset. It is established that SGD can help identify unexpected and general, size-independent, patterns within the dataset of gold cluster configurations.

## 2. Subgroup Discovery

The concepts of subgroup discovery originate from the early 90's, when the advent of large databases motivated the development of explorative and descriptive analytics tools as an interpretable complement to the supervised learning (or global modeling) paradigm [28,30,32,51-54]. Below we will start with a discussion of the three main components of subgroup discovery: (i) the use of a description language for identifying subpopulations within a given pool of data (section 2.1); (ii) the definition of utility functions that formalize the interestingness (quality) of subpopulations (section 2.2); and (iii) the design of a Monte Carlo search algorithm to find selectors that describe interesting subpopulations (section 2.3). As notational convention, we write $|X|$ to refer to the number of elements contained in $X$ (the cardinality of $X$), and $\mathcal{P}(X)$ to refer to the set of all possible subsets of $X$ (its power set). Moreover, we denote by $X \setminus Y$ the set of elements of $X$ that are not contained in $Y$ (its set difference). Logical conjunctions AND, OR, and NOT, are represented by the symbols ∧, ∨, and ¬.

## 2.1. Description Language

Consider some population $P$ of materials that contains subgroups corresponding to subpopulations $P' \subseteq P$ that exhibit yet unknown regularities with respect to some properties of interest. The population of materials is assumed to be represented by a set of features, where each feature is a map (a function) $a: P \rightarrow V$ of the population into some value domain $V$. The implicit assumption is that features can be measured and are comparable across all materials in the population. In the context of a specific analysis question, the features are partitioned into two subsets: *description features* $A$ and *target variables* $T$. The description features are used to describe subpopulations, whereas the target variables determine how important specific subpopulations are to our question being analyzed. For example, if we are interested in examining the band gaps of materials, then the target variable is the band gap and the description features are the material's properties, such as the number of atoms in a crystal unit cell, the composition, and the radii of its atomic $s$ and $p$ orbitals. For lists of features that we consider in our study, see Tables S1 and S2 in the Supporting Information.

The next aspect of our analysis is the *basic selectors*. The definition of basic selectors is an important design step that determines the interpretability and interestingness of the subgroup descriptions [32]. Basic selectors are statements regarding the features such as "the band gap of the material is large" or "the material has a rocksalt crystal structure". A set of basic selectors $C$ are constructed from the description features $a: P \rightarrow V$ according to different rules depending on their type: categorical, ordinal and metric (see Table 1 for examples). For *categorical* features, i.e., when $V$ is a discrete set with no relevant internal order, basic selectors of the form $c(p) \equiv a(p) = v$ for all values $v \in V$ are constructed. For *ordinal* features, i.e., when $V$ contains a set of discrete and ordered values, but the scale cannot be used to interpolate between values in a meaningful way, inequality constraints $c(p) \equiv a(p) \leq v$ and $c(p) \equiv a(p) \geq v$ for all $v \in V$ are used. For *metric* features, i.e., the values are from a continuous ordered scale that adheres to a meaningful notion of distance, selectors similar to the ordinal case are constructed. In this case, however, we cannot simply use all possible cut-off values, but instead have to find a small computationally feasible subset. Ideally, we would like to allow the SGD algorithm to either completely select or to completely deselect all groups of materials with very similar feature values. This goal can be approximated by finding cut-off values through $k$-means clustering [55]. That is, for a desired number $k$ of cut-off values we find a set $R \subseteq V$ of $k+1$ representative values that minimize the sum of squared differences $\sum_{p \in P}(a(p) - r_{a(p)})^2$, where $r_v$ minimizes $|v - r|$ among $r \in R$ for some $a$-value $v$. In this way, each $a$-value in the population $P$ is assigned to a cluster represented by one element in $R$. The cut-off values are then given as the arithmetic mean between the maximum and the minimum $a$-value of neighboring clusters.

**Table 1. Examples of basic statements constructed from categorical, ordinal, and metric features.**

| Categorical | The material has a rocksalt crystal structure | The gold cluster has a planar geometry |
|---|---|---|
| Ordinal | The material has $\geq N$ atoms per unit cell | The gold cluster has a mean atom coordination $\leq X$ |
| Metric[a] | The band gap of the material is *high* | The chemical hardness of the gold cluster is *low* |

[a] The notion of *high* and *low* is based on cut-off values of the metric features determined via $k$-means clustering.

Based on a final set of basic selectors $C$, subgroup descriptions are formed as complex Boolean *selectors* $\sigma: P \rightarrow \{\text{true, false}\}$ defined through conjunctions

$$\sigma(\cdot) \equiv c_1(\cdot) \wedge \ldots \wedge c_l(\cdot) \tag{1}$$

of basic statements $c_1, \ldots, c_l \in C$. Analogously to previous work by some of the authors [22], we define the *descriptor* induced by $\sigma$ as the set of descriptive features that are referenced in $\sigma$ [56]. The subpopulation of $P$ that is defined by $\sigma$ is called the *extension* of $\sigma$ and is written as

$$\text{ext}(\sigma) = \{p \in P: \sigma(p) = \text{true}\} \tag{2}$$

Although this definition yields $2^{|C|}$ possible subgroup selectors for a given set of basic selectors, usually only a few of those describe distinct and interesting subpopulations. To algorithmically determine those of interest, we have to formalize the notion of interestingness (quality) of subpopulations. As indicated above, this definition refers to the target variables $T$. In particular, let $Y = V_1 \times ... \times V_k$ denote the joint domain of all target variables. Then the utility of a selector $\sigma$ depends on the collection of $Y$ values in its extension.

## 2.2. Subgroup Quality

The SGD literature utilizes different notions of subgroup quality depending on the type and number of target variables, as well as on the kind of patterns to be discovered [32,51]. A shared characteristic between quality functions is that they are usually a weighted product of two factors corresponding to the relative size of the selected subpopulation and the utility of the selection. Formally, for a weight parameter $\alpha \in [0,1]$ we consider quality functions of the form

$$q(\sigma) = \text{cov}(\sigma)^\alpha u(\text{ext}(\sigma))^{1-\alpha} \qquad (3)$$

where $\text{cov}(\sigma) = |\text{ext}(\sigma)|/|P|$ is the *coverage* of $\sigma$ and $u: \mathcal{P}(P) \to \mathbb{R}$ is some *utility function*. The combination of these two factors is required because individually each of them is trivially maximized by either extremely general selectors (maximizing coverage) or extremely specific selectors (maximizing utility). An alternative view on considering size is that it plays a similar role as the regularization term in Ridge Regression or LASSO [57,58]. We use an $\alpha$-value of 0.5 unless mentioned otherwise, which puts equal importance on the generality and the utility of findings.

Regarding the utility function, the traditional focus of SGD is to look for subgroups that exhibit target values with a distribution that differs as much as possible from the distribution of the target variables in the global population. A representative example is the *(absolute) mean shift function* $u_m(S) = |m_S - m_P|$ for a single metric target variable, i.e., $T = \{a\}$ with $a: P \to \mathbb{R}$, where $m_U = \sum_{p \in U} a(p)/|U|$ is the mean value of $a$ in a population $U$. Although focusing on large deviation can lead to interesting findings, it has some problems in our application context of materials datasets: (1) Deviation in itself neglects *consistency* in the sense that subgroups with a high target deviation might have a poor model fit of the target variable. For example, in a heterogeneous materials-science data set a subgroup might have a large mean shift but have a local standard deviation that is higher than the global standard deviation. This is problematic for our goal of uncovering physical relations between material structure and properties, for which we want our findings to be highly consistent. (2) Focusing on deviation has the implicit assumption that the global reference distribution is already well understood and distance from it is therefore meaningful. However, the global distribution of properties in big-data of materials is often an effect of a large number of mixed factors and therefore often too complex to describe in a compact way – in fact this complexity is one of the reasons to resort to local modelling via subgroup discovery in the first place.

Therefore, the utility functions we define for this study aim for consistent findings by formalizing different notions of purity in the distribution of target values. In particular, we consider the following utility functions:

- The *(normalized) information gain* $u_{ig}(S) = (H_P - H_S)/H_P$ for categorical target variables $T = \{a_1, ..., a_k\}$ with joint domain $Y$, where $H_U = -\sum_{y \in Y} \pi_U(y) \log \pi_U(y)$ is the Shannon entropy [59] of the empirical probabilities $\pi_U(y) = |\{p \in U: (a_1(p), ..., a_k(p)) = y\}|/|U|$ (defining $\pi_U(y) \log \pi_U(y) = 0$ for $\pi_U(y) = 0$). This measure is maximized by populations with point distributions of target values (i.e., such that there is a $y \in Y$ with $\pi_U(y) = 1$) and minimized by those that have a uniform target distribution.

- The *(standard) variation reduction* $u_{vr}(S) = (s_P - s_S)/s_P$ where

$$s_U = \sqrt{\sum_{p \in U} \frac{(m_U - t(p))^2}{|U| - 1}} \qquad (4)$$

is the sample standard deviation in the case of a single metric target variable $T = \{a\}$, $a: P \to \mathbb{R}$, with empirical mean $m_U = \sum_{p \in U} a(p)/|U|$ (and in the case of multiple metric target variables, the squared differences can for example be replaced by the squared norm of the difference vectors between sample values and the mean vector). Similar to the information gain $u_{ig}$, this utility function favors subgroups where the target values are as close as possible to some localized value over groups with uniform distribution, only this time the deviation from a point distribution is measured in a metric sense.

- The *(Pearson) correlation gain* $u_{cg}(S) = (|r_S| - |r_P|)/(1 - |r_P|)$ between pairs of numeric target variables $T = \{a, b\}$, $a, b: P \rightarrow \mathbb{R}$, where

$$r_U = \frac{1}{|U| - 1} \sum_{p \in U} \left(\frac{m_U^a - a(p)}{s_U^a}\right) \left(\frac{m_U^b - b(p)}{s_U^b}\right) \tag{5}$$

is the sample Pearson product-moment correlation coefficient of the paired $a$ and $b$-values in the population $U$ (with $m_U^x$ and $s_U^x$ being the sample mean and the standard deviation of target variable $x$ as in the definition of the variation reduction utility function). This utility function is maximized (having a value of 1) for subpopulations where the paired target values all lie on a line ($|r_S| = 1$). Hence, it is used to find subgroups where there is an approximately linear relationship between two metric target variables. This is in contrast to traditional variants where subgroups with an unusual correlation (e.g., inverse effects) are sought [28].

The usage of these utility functions for uncovering interpretable local patterns and descriptors is demonstrated in Section 3. Beforehand, we describe a simple and robust algorithm for finding subgroups with high quality values.

### 2.3. Search Strategy

Optimizing any of the above quality functions is a computationally hard problem with no known polynomial time approximation algorithm (note that the size of the search space has an exponential relation to the number of basic selectors considered). The standard algorithmic approach to find optimal subgroups are exponential time branch-and-bound algorithms, which can be effective for certain input datasets if a good bounding function for the employed quality function is known [60]. Although deriving such bounding functions is an interesting research problem, here we follow a different route and utilize a heuristic two-step Monte Carlo sampling approach, which works well for many practical problems [61,62]. The following procedure is repeated iteratively for as many random result patterns as desired:

1. *Seed generation*: Sample a random seed conjunction $\sigma_0$ with generation probability $\mathbb{P}(\sigma_0 = \sigma)$ proportional to the size of the extension $|\text{ext}(\sigma)|$. This can be implemented by a direct sampling approach in time $O(mk)$, where $m$ denotes the number of data points and $k = |C|$ the number of basic selector functions [62]. The idea is to first sample a member of the global population with a probability proportional to the number of conjunctions of basic selectors that are true for that population member, and then to sample uniformly a selector from those conjunctions.
2. *Opportunistic pruning*: Starting from $\sigma_0(\cdot) \equiv c_1(\cdot) \wedge \ldots \wedge c_l(\cdot)$, with the basic selectors in random order, consider each $c_i$ for $i \in \{1, \ldots, l\}$ and remove it if the quality $q(\sigma'_{i-1}) \geq q(\sigma_{i-1})$, where $\sigma'_{i-1}$ results from $\sigma_{i-1}$ by removing $c_i$ (in this case define $\sigma_i = \sigma'_{i-1}$, otherwise $\sigma_i = \sigma_{i-1}$). Since all quality functions considered here can be computed in time $O(m)$, the worst-case time complexity of this step is $O(mk)$.

For our analysis, subgroup selectors are chosen based on having the highest value of the quality function. At least 10 000 random seeds are used in the Monte Carlo search of subgroup selectors. Upon reapplication of the Monte Carlo algorithm, the same optimal subgroup selectors are found for the patterns described below. Nevertheless, due to its non-exhaustive nature, the Monte Carlo procedure does not guarantee that superior unfound patterns do not exist. The SGD algorithm was implemented in the Creedo web application with the realKD library [63].

## 3. Application

### 3.1. Octet Binary Semiconductors: Toward Predicting Crystal Structures

Predicting the crystal structure of a material from only knowledge of its chemical composition is a long-standing goal to facilitate the design of materials [2,64,65]. Over four decades ago, van Vechten and Phillips [35,36,66] analyzed the classification of octet binary (OB) semiconductors and proposed a descriptor to classify the zincblende (ZB), wurtzite (WZ) and rocksalt (RS) structures. Since the studies by Van Vechten and Phillips, many researchers have sought to identify superior descriptors to classify the OB compounds (see Refs. [22,39,40]

and the references within). Descriptors used to classify the OB crystal structures were typically introduced based on understanding by the authors of the bonding nature of these materials.

As a more general and less biased approach, recently a two-step feature selection algorithm, which consists of using the least absolute shrinkage and selection operator method followed by $\ell_0$-constrained minimization (LASSO+$\ell_0$), was utilized to systematically discover an interpretable descriptor that provides a classification and even quantitative energy differences of the OB compounds (as either ZB or RS) [22]. Figure 2 shows the main conclusion. The two-dimensional descriptor is derived from solely combinations of the radii of the maximum probability density of the valence $s$ ($r_s$) and valence $p$ ($r_p$) orbitals of the free atoms A and B that make up the octet binary compounds, as well as their ionization potential (IP) and electron affinity (EA).

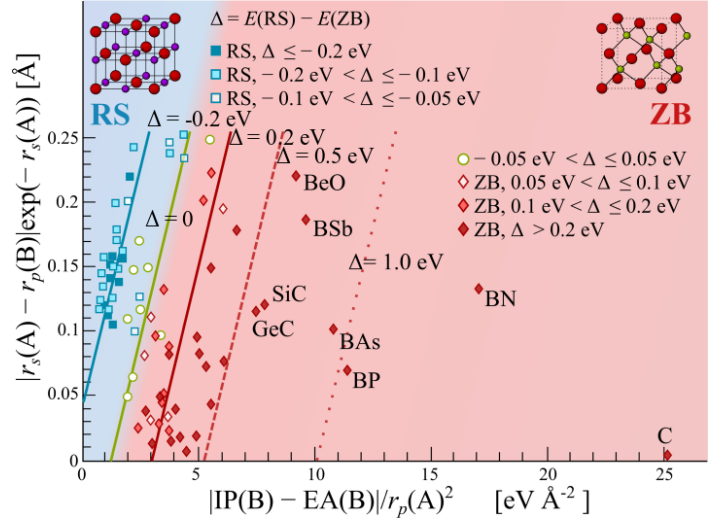

**Figure 2.** Computed energy differences ($\Delta$) between rocksalt and zincblende crystal structures of the 82 octet AB-type binary compounds organized according to the two-dimensional descriptor found using LASSO+$\ell_0$ [22].

As a complementary local modeling approach to the LASSO+$\ell_0$ global modeling paradigm, here we examine the crystal-structure classification of octet binary semiconductors using subgroup discovery. The dataset of the OB compounds is obtained from Ghiringhelli *et al.* [22] (the full list of the OB compounds is provided in their Supporting Information and all the input and output files can be downloaded from the NOMAD Repository using http://link.aps.org/doi/10.1103/PhysRevLett.114.105503 as an external DOI reference). The feature space is restricted to atomic properties of the free atoms within the OB compounds; in total, 55 features are used in the subgroup discovery algorithm (crystal structure information, 14 atomic properties, and 40 features derived from those; see Table S1 of the Supporting Information). For the application of SGD, the information gain utility function $u_{\text{ig}}$ is used with the sign of the energy difference between RS and ZB as the target variable ($T = \{\text{sign}(\Delta)\}$). The difference in energy between ZB and WZ structures for these materials is very small: the maximum absolute difference is 0.04 eV and the average over the dataset is 0.01 eV. Therefore, as in Ref. [22], we do not distinguish between ZB and WZ and we use the energy of ZB as the reference for $\Delta$. A set $C$ of 1 576 basic selectors is generated from the remaining 54 description features according to the rules outlined in Section 2.1.

Application of SGD identifies several pure subgroups that exclusively contain either only RS or only ZB structures. Figure 3a shows the simplest (smallest number of basic selectors) maximum quality selectors found for each of the two crystal structures. The RS subgroup and ZB subgroup are described by the selectors $\sigma_1^{\text{RS}} \equiv |r_p^A - r_p^B| \geq 0.91 \text{ Å} \wedge r_s^A \geq 1.22 \text{ Å}$ and $\sigma_1^{\text{ZB}} \equiv |r_p^A - r_p^B| \leq 1.16 \text{ Å} \wedge r_s^A \leq 1.27 \text{ Å}$, respectively. Here superscripts A and B denote the free atoms that make up the octet AB-type semiconductors. Since both subgroups are pure, they have the maximum entropy gain ($u_{\text{ig}}(\sigma_1^{\text{RS}}) = u_{\text{ig}}(\sigma_1^{\text{ZB}}) = 1$, where for notational convenience we write $u(\sigma)$ for $u(\text{ext}(\sigma))$). The RS subgroup covers all but two compounds of its respective crystal structure group ($\text{cov}(\sigma_1^{RS}) = 39/82$), whereas the ZB subgroup covers all but one ($\text{cov}(\sigma_1^{ZB}) = 40/82$).

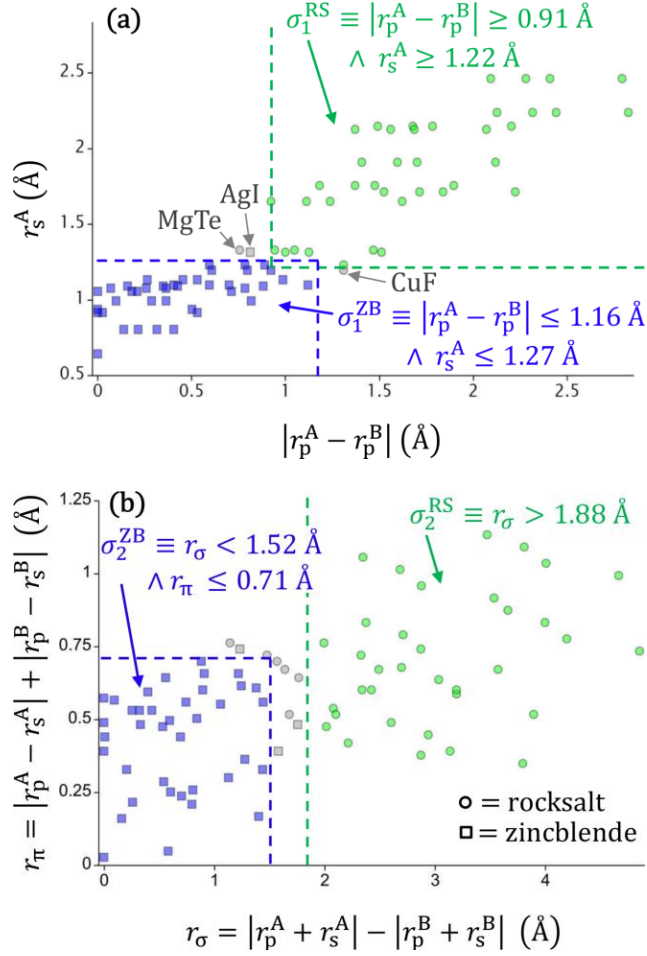

**Figure 3.** Application of subgroup discovery to the 82 octet binary semiconductors helps us identify interpretable selectors that describe subgroups of the rocksalt (RS) and zincblende (ZB) structures. (a) The subgroups described by selectors $\sigma_1^{RS}$ (describing the RS subgroup) and $\sigma_1^{ZB}$ (describing the ZB subgroup) with the highest quality value for a two-dimensional descriptor. (b) The subgroups described by selectors consisting of St. John and Bloch's $r_\sigma$ and $r_\pi$ descriptors [33]. The dashed blue and green lines denote the (non-linear) intersection of axis-parallel hyperplanes that contain the ZB and RS subgroups. The squares and circles denote zincblende and rocksalt crystal structures, respectively. Green: rocksalt subgroup described by $\sigma_i^{RS}$; Blue: zincblende subgroup described by $\sigma_i^{ZB}$; Grey: points described by neither selector. Another representation of $\sigma_1^{RS}$ and $\sigma_1^{ZB}$ is shown in Figure S1 of the Supporting Information, where instead the axes are chosen to be the two-dimensional descriptor found by Ghiringhelli and coworkers using LASSO+$\ell_0$ [22].

A global model can be created by combining the underlying descriptors of both subgroups shown in Figure 3a, i.e., the descriptive features that are referenced in both $\sigma_1^{RS}$ and $\sigma_1^{ZB}$, which describe 79 of the 82 octet binary compounds correctly as either RS or ZB – and that is agnostic about only three structures that have a nearly degenerate energy difference between RS and ZB (36.5 meV for AgI, 19.0 meV for CuF, and 4.5 meV for MgTe). This two-dimensional descriptor consists of only linear combinations of atomic radii of the valence *s* and *p* orbitals, i.e., $\{r_s^A, |r_p^A - r_p^B|\}$. SGD helps find that OB semiconductors with relatively larger values of $r_s^A$ and $|r_p^A - r_p^B|$ favor RS structures, whereas smaller values favor ZB structures. Large valence *p* radii differences between atomic elements suggests ionic character within compounds, whereas smaller atomic radii differences suggest covalent character, which is in agreement with reports that ionic OB compounds typically form RS structures [35,36,39,66].

Figure 3b depicts RS and ZB subgroups described by selectors consisting of previously reported descriptors; namely, St. John and Bloch's $r_\sigma$ and $r_\pi$ descriptors are used, which approximate the *s-p* contribution to the electronegativity and the *s-p* hybridization [33]. Interestingly, the descriptors that make up the optimally found subgroup selectors (Figure 3a) are similar to $r_\pi$ and $r_\sigma$ (Figure 3b). Selectors using $r_\sigma$ and $r_\pi$ as descriptive features describe 38 of the 41 ZB structures and 35 of the 41 RS structures. On the other hand, subgroups described by selectors that consist of the two-dimensional descriptor found by Ghiringhelli *et al*. using LASSO+$\ell_0$ contain only 71 of the RS and ZB structures (see Figure S2 of the Supporting Information). This illustrates the different modeling strategies used by LASSO+$\ell_0$ and subgroup discovery, i.e., LASSO+$\ell_0$ optimizes for a single linear separating hyperplane, whereas subgroup discovery uses a (non-linear) intersection of axis-parallel hyperplanes. The LASSO+$\ell_0$ algorithm is suited to find a globally optimal and sparse descriptor for crystal-structure classification [22], which is complementary to the subgroup discovery methodology. However, subgroup discovery could find multiple local models that, when combined, span a large and relevant portion of the dataset. This could be a useful strategy to build structure maps for crystal-structure prediction [65].

### 3.2. Neutral Gas-Phase Gold Clusters: The Search for Structure-Property Relationships

Subgroup discovery is applied here to ascertain interpretable and general patterns between physicochemical and geometrical properties for 24 400 neutral gas-phase gold clusters (sizes of 5-14 atoms). Gold clusters have been a topic of sustained interest due to their various important and unique electronic, optical, and catalytic properties [67-75]. However, the majority of past computational studies on such clusters focused on a static, monostructure, description at 0 K, but increasing amounts of evidence indicate dynamic structural disorder is a common feature among clusters and that numerous isomers can coexist at finite temperature. General patterns holding across multiple cluster isomers of various sizes at finite temperature may be missed by this standard approach, therefore below we analyze gold clusters generated from the canonical ensemble at various temperatures.

*Ab initio* replica-exchange molecular dynamics (REMD) [76] simulations are performed with the FHI-aims electronic-structure code [77] to generate the gold cluster configurations based on uniform sampling of the canonical ensemble at temperatures from 100 to 814 K. REMD simulations utilized *light-tier 1* (excluding the hydrogenic 6*h* basis functions) numerical settings with energies and forces obtained from spin-polarized DFT with the PBE exchange-correlation functional [78] corrected for many-body dispersion (MBD) [79] (which we denote as PBE+MBD). Relativistic effects are treated using the 'atomic ZORA' scalar correction [80,81]. See Ref. [75] for validation of the used REMD [82] settings and the choice of exchange-correlation functional, as well as the importance of including van der Waals corrections. Gold cluster geometries and their features are sampled from each replica in the REMD simulation every 1.0-2.3 ps, yielding 2 440 configurations per gold cluster size (24 400 configurations total for $Au_5$-$Au_{14}$). All patterns identified using subgroup discovery were preserved upon resampling of the gold cluster configurations.

The features computed for each cluster geometry are: relative total energy $\Delta E$ (relative to the most stable structure at each size), normalized radius of gyration $R_{g0}$, ionization potential IP, electron affinity EA, HOMO-LUMO energy gap $E_{HL}$, cluster size $N$, replica temperature $T$, atom coordination histogram (a vector containing the number of atoms with a certain bond coordination number) [83], and relative intramolecular van der Waals energy $\Delta E_{vdW}$ (relative to the structure with the largest van der Waals energy at each size), among others. A list of the gold cluster features is provided in Table S2 of the Supporting Information and all the configurations can be downloaded from the NOMAD Repository (http://dx.doi.org/10.17172/NOMAD/2016.11.02-1). Since at finite temperatures a gold cluster will never be exactly at a minimum on the potential energy surface, all features of the gold clusters are computed from the unrelaxed structures generated from the canonical ensemble.

*3.2.1. Finding patterns of the HOMO-LUMO energy gap*

The HOMO-LUMO energy gap ($E_{HL}$) of neutral gold clusters is known to oscillate depending on whether the cluster has an even or odd number of atoms [84,85]. The even-odd oscillatory behavior of $E_{HL}$ is due to spin pairing for the even-sized neutral gold clusters and the lack of spin pairing for the odd-sized clusters. As a tutorial example, we first demonstrate that the SGD algorithm can help rediscover the known phenomenon that neutral clusters with an odd number of atoms have a small $E_{HL}$ relative to neutral clusters with an even number of atoms. Algorithmically, subgroups of gold cluster configurations with a low standard deviation of the HOMO-LUMO energy gap are sought. That is, the target variable is specified as $T = \{E_{HL}\}$ and the standard variation reduction utility function $u_{vr}$ is used (see section 2.2 for the definition of $u_{vr}$). A set $C$ of 338 basic selectors is generated (using 6 cut-off values for metric variables). Examples of basic selectors are statements such as "the number of atoms in the gold cluster is an even number", "the number of the atoms in the gold cluster is $\geq 8$," or "the energy of the gold cluster configuration is low (i.e., an energetically favorable configuration)."

Upon application, SGD finds that the highest quality subgroup is described by the selector $\sigma_1^{HL} \equiv \text{odd}(N)$, which has a coverage of $\text{cov}(\sigma_1^{HL}) = 0.50$ (the subgroup covers 50% of the total population) and a utility value of $u_{vr}(\sigma_1^{HL}) = 0.89$ (89 percent of the standard deviation of the HOMO-LUMO energy gap is reduced in this subgroup relative to the global data). In other words, the clusters with an odd number of atoms ($N$ = 5, 7, 9, 11 and 13) form a well-defined subgroup, Figure 4a. The second highest quality subgroup found is described by $\sigma_2^{HL} \equiv \text{even}(N) \wedge N \geq 7$, which has $\text{cov}(\sigma_2^{HL}) = 0.40$ and $u_{vr}(\sigma_2^{HL}) = 0.61$. Interestingly, Au$_6$ is excluded from this subgroup because it has an unusually large $E_{HL}$ due to its highly stable ground state structure as a result of its σ-aromaticity [46,85,86]. As shown in Figure 4b, the distribution of the HOMO-LUMO energy gap values indicates three clear subgroups; namely, $\text{odd}(N)$, $\text{even}(N) \wedge N \geq 7$, and $N = 6$. In Figure 4c, the oscillation in the HOMO-LUMO gap with respect to cluster size is observed, as well as the fact that larger even-sized gold clusters generally have a decreasing average HOMO-LUMO gap (due to their increasingly metallic nature). Moreover, the HOMO-LUMO energy gap varies dramatically depending on cluster configuration at a given size.

*3.2.2. Structural and electronic properties of planar/quasi-planar and nonplanar gold clusters*

Subgroup discovery is next applied to discern general patterns among the structural and electronic features of planar/quasi-planar and nonplanar (compact, three-dimensional) structures. Small gold clusters often adopt stable planar geometries as a consequence of relativistic effects [87,88], and planar and nonplanar structures can coexist simultaneously at finite temperature [89-91]. It is important to emphasize that theoretical studies using traditional generalized gradient functionals (without van der Waals corrections) are biased towards planar structures for gold clusters [47,49]; however, our benchmark studies using PBE+MBD with *tight-tier 2* settings have reasonable agreement with isomer energetics predictions using HSE06+MBD as well as RPA@PBE. Geometries and energy differences between the lowest energy planar and nonplanar clusters using PBE+MBD are provided in Figure S3 of the Supporting information. Detailed analysis of the choice of exchange-correlation functional on relative isomer stabilities between planar and nonplanar neutral gold clusters as well as the importance of temperature on cluster isomer probabilities based on analysis of free energy surfaces will be reported in a separate article.

The minimum thickness of a gold cluster configuration has been used as an order parameter to monitor the planarity of a cluster [89]. Instead, here the planar/quasi-planar (from here on called planar or 2D) and nonplanar (3D) gold clusters are approximately discriminated based on their normalized radius of gyration. The radius of gyration is computed as the root mean square distance of the cluster's parts from its center of mass. The normalized radius of gyration is obtained by dividing the radius of gyration of each cluster configuration by the radius of gyration of the lowest energy planar structure at each size. Planar clusters have a larger radius of gyration compared with nonplanar structures due to their less compact nature. Cut-off values for discriminating between planar and nonplanar clusters and are chosen based on examining the probability distribution of the radius of gyration (See Table S3 and Figure S4 of the Supporting information). We refer to this categorical feature as "Shape", i.e., planar or nonplanar.

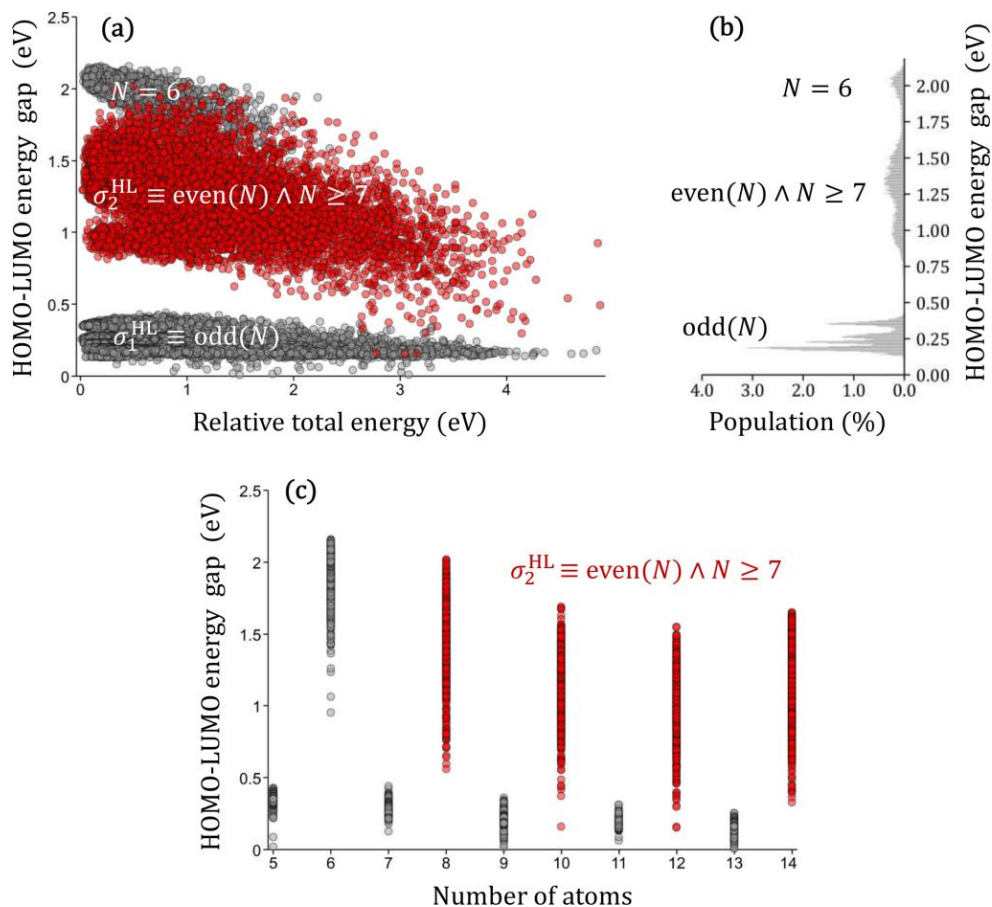

**Figure 4.** Gold cluster configurations are examined for subgroups of the HOMO-LUMO energy gap with a low standard deviation. (a) The HOMO-LUMO energy gap of each cluster is shown against its relative total energy (i.e., the energy of each cluster configuration is referenced against its most stable structure for each size). (b) The population of gold clusters with a certain value the HOMO-LUMO energy gap, with the three main subgroups labeled. (c) The HOMO-LUMO energy gap of each cluster configuration as a function of cluster size. Red: subgroup described by $\sigma_2^{HL}$.

Although there is a broad distribution of HOMO-LUMO energy gaps depending on cluster size and geometry, no statistically significant local patterns between $E_{HL}$ and planar or nonplanar structures are found using SGD. The Monte Carlo procedure does not guarantee that unfound patterns do not exist, and thus we cannot ascribe significance to the lack of a found pattern. However, application of SGD using the information gain utility function $u_{ig}$ with planar or nonplanar categorization as the target variable $T = \{Shape\}$ elucidates that the majority of low energy (stable) planar structures have an average atom coordination number of less than or equal to three (see Figure 5). The identified subgroup selector is $\sigma_1^{Sh} \equiv \Delta E \leq 1.88$ eV $\wedge$ #mean $\leq 3$. The extension $S = \text{ext}(\sigma_1^{Sh})$ described by $\sigma_1^{Sh}$ contains 14 097 configurations (coverage of 58%) of sizes 5-14 atoms that are almost exclusively planar ($\pi_S(2D) = 0.98$ and $u_{ig}(\sigma_1^{Sh}) = 0.86$). In other words, low-energy planar clusters with 5 to 14 atoms typically have a coordination number less than or equal to three (although the subgroup is dominated by clusters of $Au_5$-$Au_{10}$). The high coverage of this subgroup reinforces the notion that gold cluster isomers within this size range typically adopt planar geometries due to their energetic stability [49]. Note, $Au_{13}$ and $Au_{14}$ rarely adopt planar gold cluster configurations [49,91] at finite temperature and therefore are not widely included in this subgroup (Figure 5b).

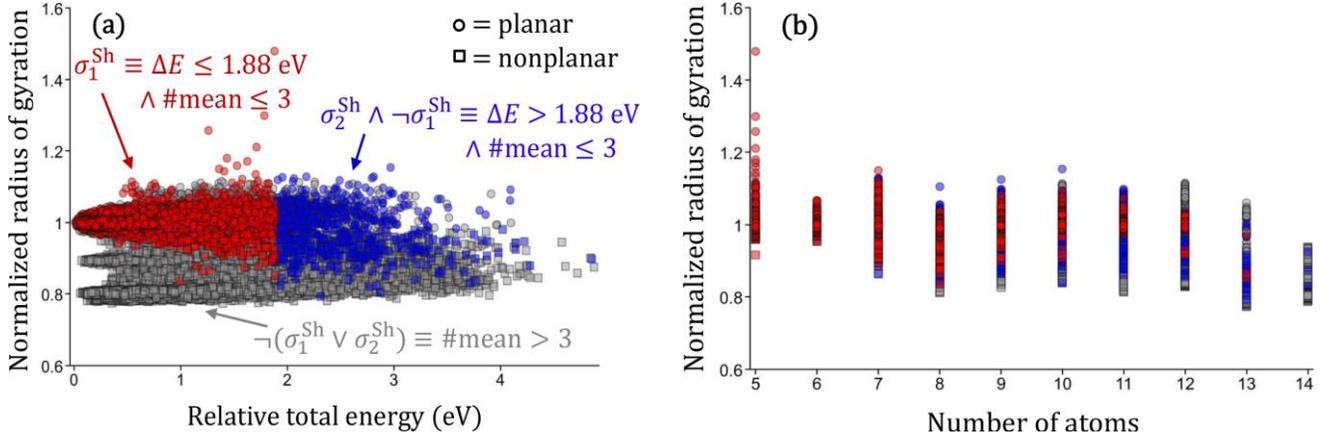

**Figure 5.** Subgroup selectors are identified that describe the structural and electronic features of planar/quasi-planar structures. (a) The normalized radius of gyration of each cluster configuration is shown against its relative total energy. (b) The normalized radius of gyration of each cluster configuration as a function of size. Red: subgroup described by $\sigma_1^{Sh}$; Blue: additional points included in generalized variant $\text{ext}(\sigma_2^{Sh} \wedge \neg \sigma_1^{Sh}) = \text{ext}(\sigma_2^{Sh}) \setminus \text{ext}(\sigma_1^{Sh})$; Grey: points described by neither selector $\neg(\sigma_1^{Sh} \vee \sigma_2^{Sh}) = \neg \sigma_2^{Sh}$.

One can examine the importance of the individual descriptive features making up the selectors by removing them and examining the generalized selector's properties. For example, if the descriptive feature $\Delta E \leq 1.88$ eV is removed from $\sigma_1^{Sh}$ then the generalized selector $\sigma_2^{Sh} \equiv \#\text{mean} \leq 3$ is produced, which describes planar structures with less purity ($u_{ig}(\sigma_2^{Sh}) = 0.77$). It is unexpected that the average atom coordination number of low-energy planar gold clusters remains $\leq 3$ across a broad range of clusters sizes; for an infinite fcc(111) the average atom coordination is six, and the "inner" atoms of planar gold clusters are also six-fold coordinated. In some cases, the low average atom-coordination number of planar clusters relative to their more compact, nonplanar, isomer counterparts may result in increased reactivity due to the presence of more under-coordinated sites and better electron-accepting capabilities [92-94].

### 3.2.3. Intra-cluster van der Waals interactions and its shape dependence

Van der Waals (vdW) interactions are critical to include in electronic-structure theory calculations for determining the correct energetic ordering of material configurations, e.g., adsorption configurations of molecules on surfaces [95] and peptide conformers [96]. It has been suggested that nonplanar cluster configurations are more stabilized by intramolecular vdW interactions than planar cluster configurations [91], but this has yet to be quantified or demonstrated across a large range of cluster sizes and geometries at finite temperature. To test this hypothesis, SGD is applied using the variation reduction utility function $u_{vr}$ with the intramolecular vdW energy (referenced to the maximum at each size) as the target variable ($T = \{\Delta E_{vdW}\}$), Figure 6a. Note, recall here that $\Delta E_{vdW}$ is just the long-range correlation energy within the many-body dispersion framework [79]. The highest quality selector found is $\sigma_1^{vdW} \equiv N \geq 8 \wedge N \leq 10 \wedge \text{Shape}(2D)$ with $\text{cov}(\sigma_1^{vdW}) = 0.26$ and $u_{vr}(\sigma_1^{vdW}) = 0.84$. This subgroup describes the phenomena that planar (2D) gold clusters generally exhibit weaker vdW interactions than nonplanar clusters for $Au_8$-$Au_{10}$ clusters. Although the $Au_8$-$Au_{10}$ pattern is highlighted in Figure 6a, subgroups describing similar phenomena for sizes 11-12 are also found.

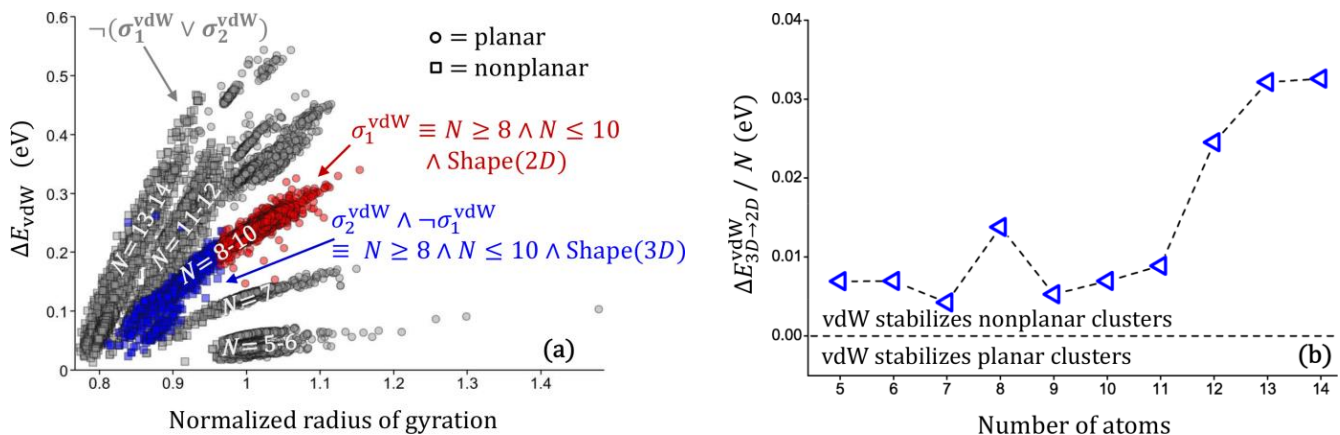

**Figure 6.** Gold cluster configurations are examined for patterns involving intra-cluster van der Waals interactions. (a) SGD finds subgroup selectors using the variation reduction utility function with the relative intramolecular vdW interaction energy $\Delta E_{vdW}$ (referenced to its maximum value at each size) as the target variable. Red: subgroup described by $\sigma_1^{vdW}$; Blue: additional points included in generalized variant $\text{ext}(\sigma_2^{vdW} \wedge \neg \sigma_1^{vdW}) = \text{ext}(\sigma_2^{vdW}) \setminus \text{ext}(\sigma_1^{vdW})$; Grey: points described by neither selector $\neg(\sigma_1^{vdW} \vee \sigma_2^{vdW}) = \neg \sigma_2^{vdW}$. (b) The intramolecular vdW energy difference per atom between the lowest energy planar and nonplanar gold cluster geometries as a function of size (based on structures in Figure S3 of the Supporting Information).

The pattern found by SGD is further supported based on examination of the vdW energy difference between the lowest energy planar and nonplanar structures as a function of size. The vdW interaction per atom becomes increasingly larger in nonplanar clusters relative to planar clusters as the cluster size increases, Figure 6b. Additionally, the visualization in Figure 6a suggests the existence of an unexpected higher order pattern: for certain cluster sizes there appears to be a linear relationship between the normalized radius of gyration and the vdW energy. Subgroup discovery is used to test this hypothesis by augmenting the set of target variables to $T' = \{R_{g0}, \Delta E_{vdW}\}$ and using the correlation gain utility function $u_{cg}$. SGD finds $\sigma_3^{vdW} \equiv N \geq 8$ with $\text{cov}(\sigma_3^{vdW}) = 0.70$ and a correlation gain of $u_{cg}(\sigma_3^{vdW}) = 0.49$, which corresponds to a local Pearson product-moment correlation coefficient of $r_s = 0.84$. By modifying the weight parameter $\alpha$, smaller groups with an even more linear relationship can be found, e.g., $\sigma_4^{vdW} \equiv N \geq 13$ with $r_s = 0.97$ using $\alpha = 1/8$. The compact nature of the nonplanar gold clusters increases its intra-cluster vdW interactions relative to planar clusters, even though the nonplanar clusters are less polarizable [97]. The strong influence of vdW interactions on stabilizing nonplanar gold structures relative to planar structures suggests that an accurate treatment of vdW interactions is required for predicting the correct isomer energetic ordering of polarizable nanoclusters.

*3.2.4. Analyzing relationships between chemical hardness and cluster stability*

The concept of chemical hardness is typically understood as the resistance of a system's chemical potential to a change in the number of electrons, and thus it is often used as a reactivity index [98-101]. Correlations have been found between the chemical hardness, stability, polarizability, and size of different systems [102-104]. Statistical mechanics in the grand canonical ensemble suggests that the ground state structure of a system has the maximum hardness of all the possible states at 0 K [105-107]. As a manifestation of the principle of maximum hardness, relatively more stable lithium clusters (those having a magic number of atoms) were predicted to have a local maximum in their chemical hardness [104]. However, to what degree correlations between chemical hardness and stability are present in a large set of cluster configurations in the canonical ensemble is not known.

To formalize this question for SGD the target variables are set to the relative total energy (referenced to the lowest energy cluster at each size) and the chemical hardness (referenced to the maximum hardness cluster at each size) $T = \{\Delta E, \Delta \eta\}$ and the utility function to the correlation gain $u_{cg}$. The chemical hardness is calculated at 0 K, which is sufficient because thermal corrections to the hardness are small even above room temperature [108,109]. Although the global linear correlation between cluster stability and chemical hardness is small ($r_P = -0.27$),

SGD finds a selector that describes a strong local linear trend, Figure 7. The highest quality subgroup selector identified is $\sigma_1^{hd} \equiv \Delta E_{vdW} \leq 0.178$ eV $\wedge$ even$(N) \wedge$ #mode $\leq 5$, where #mode is the mode of the distribution of the atom coordination number. This subgroup has a coverage of cov$(\sigma_1^{hd}) = 0.20$ (20% of the total population) and correlation gain of $u_{cg}(\sigma_1^{hd}) = 0.54$ corresponding to a local Pearson correlation coefficient of $r_S = -0.81$.

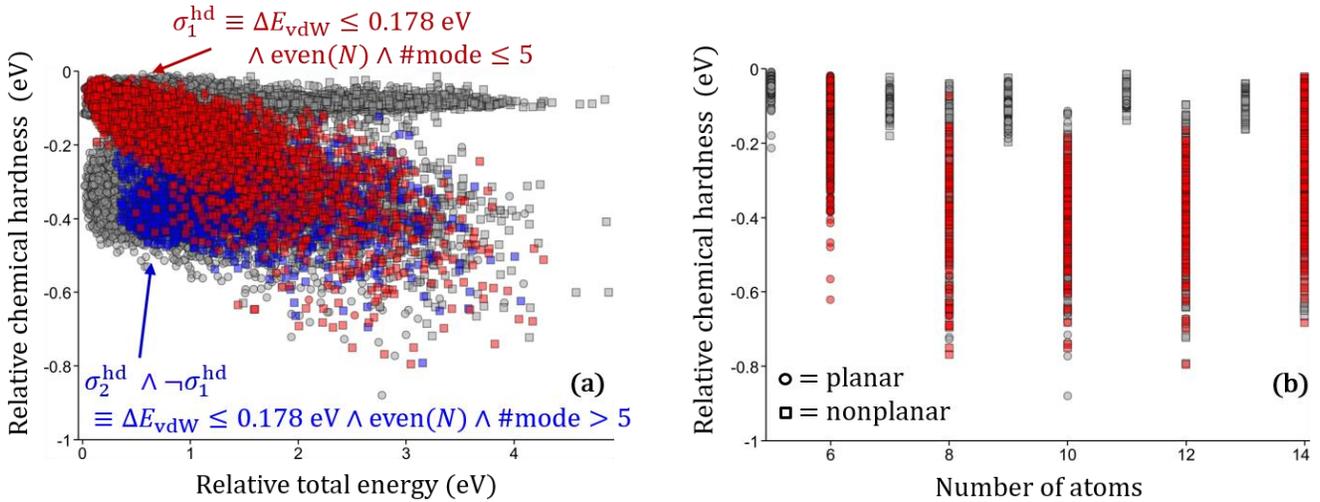

**Figure 7.** Gold cluster configurations are examined for linear correlations between the chemical hardness and the stability of the cluster. (a) The relative chemical hardness of each cluster is shown against its relative total energy. The selector $\sigma_1^{hd}$ describes a subgroup with a local linear correlation between chemical hardness and cluster stability. (b) The relative chemical hardness of each cluster configuration as a function of cluster size. Red: subgroup described by $\sigma_1^{hd}$; Blue: additional points included in generalized variant ext$(\sigma_2^{hd} \wedge \neg\sigma_1^{hd}) =$ ext$(\sigma_2^{hd}) \setminus$ ext$(\sigma_1^{hd})$. The points of ext$(\sigma_2^{hd}) \setminus$ ext$(\sigma_1^{hd})$ are displayed in Figure 7a only; Grey: points described by neither selector $\neg(\sigma_1^{hd} \vee \sigma_2^{hd}) = \neg\sigma_2^{hd}$.

It appears that the chemical hardness cannot generally predict the correct trend for the stability of multiple isomers [110]. However, within the subgroup described by $\sigma_1^{hd}$ the hardness principle qualitatively holds (with coverage of 40% of the even-sized gold clusters in the dataset). The selector $\sigma_1^{hd}$ is quite complex and relates seemingly unrelated features, i.e., van der Waals energy, atom size, and coordination number, to find a local descriptor that linearly correlates chemical hardness and stability, i.e., less stable cluster geometries are less chemically hard. Clusters with an odd number of atoms are excluded from the subgroup description because their hardness is nearly constant with geometry, which results from their unpaired electron. The importance of the constraint on the atom coordination mode can be illustrated by considering the non-optimal selector $\sigma_2^{hd}$ obtained by removing #mode $\leq 5$ from $\sigma_1^{hd}$. This subgroup described by $\sigma_2^{hd}$ has a substantially weaker linear relationship between chemical hardness and energy ($r_S = -0.66$). Further investigation into other systems for trends between chemical hardness, reactivity, and stability is required for a deeper understanding.

## 4. Conclusions

Advances in big-data analytics, i.e., statistical and machine learning, compressed sensing, and data mining, alongside the exponential growth of materials-science repositories are opening innovative avenues for identifying advanced functional materials for use in applications such as batteries, thermoelectrics, and superconductors. Nevertheless, it remains challenging to screen databases of hypothetical and known materials for anomalies and to predict materials with superior properties than existing ones. Additionally, finding predictive descriptors of materials from high-dimensional data manually is laborious, error-prone, and typically subjective. Consequently, the development and application of sophisticated big-data analytics tools to extract materials insights is required.

One promising approach is to use subgroup discovery (SGD), which is a descriptive data-mining technique to find interpretable local patterns, correlations, and descriptors of properties according to some target property (or properties) of interest. As a complement to global modeling techniques, here subgroup discovery is formulated in the context of materials science, and two exemplary systems are examined: (1) 82 octet binary semiconductors to find physically meaningful rule-based models that predict their crystal structure as either zincblende or rocksalt (RS); and (2) 24 400 configurations of neutral gas-phase gold clusters with 5 to 14 atoms to search for general structure-property relationships holding across numerous gold cluster isomers of various sizes.

In this paper, subgroup discovery is demonstrated to find an interpretable two-dimensional model consisting only of atomic radii of valence $s$ and $p$ orbitals that properly classifies 79 of the 82 octet binary structures as either rocksalt or zincblende. Since the octets are only part of over 550 known $AB_n$ binary solids [38], SGD can likely be used to find descriptors in the broader class of binary solids as well as construct structure maps, and these are directions for further investigation. For the gold clusters, unexpected and general trends are found upon application of SGD. For example, the intramolecular van der Waals interactions within planar clusters are typically significantly weaker compared with nonplanar, compact, clusters. This suggests that van der Waals interactions can be critical for accurately predicting the isomer energetic ordering of gold nanoclusters, especially for the planar to nonplanar geometrical transition.

Data analytics tools applied to materials-science data will continue to facilitate the understanding of structure-property relationships and the rational design of advanced materials. Nonetheless, limitations in both machine learning and data-mining approaches remain. In particular, for subgroup discovery an important open problem is the design of efficient optimal solvers for the quality function variants proposed in this work. Although the Monte Carlo algorithm discovers interesting patterns, it remains heuristic in nature. This prevents us to draw conclusions from the absence of certain patterns in the result set. For instance, are there hitherto undiscovered relations between the HOMO-LUMO energy gap, gold cluster stability, and gold cluster geometry? The design of optimal solvers for the proposed variants of SGD is currently underway to address this question, among others. We posit that subgroup discovery will serve as a useful tool for the extraction of insights from big data of materials, and its continued development will help pave the way toward novel materials discovery.

## Acknowledgements

The project received funding from the European Union's Horizon 2020 research and innovation program under grant agreement no. 676580 with The Novel Materials Discovery (NOMAD) Laboratory, a European Center of Excellence. BRG acknowledges support from the Alexander von Humboldt-Foundation with a Postdoctoral Fellowship. The authors thank Christopher Sutton, Matthias Rupp and Runhai Ouyang for stimulating discussions and for providing feedback on the manuscript.

## Notes

All the examples reported in this paper can be run via web-based tutorials accessible via: https://analytics-toolkit.nomad-coe.eu/Creedo/index.htm, where the users can also interactively change the input settings and compare the outcome with our results. These tutorials are part of the NOMAD analytics toolkit (https://analytics-toolkit.nomad-coe.eu/ ), which is developed in the context of the NOMAD Laboratory (https://nomad-coe.eu/ ) ⱡ These authors have contributed equally to this manuscript.

TOC IMAGE

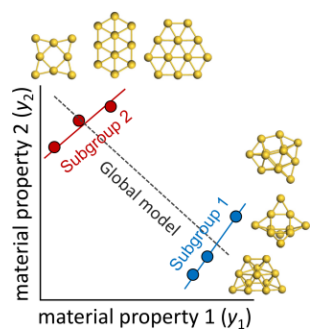